\documentclass{article}
\usepackage{graphicx}
\usepackage{times}
\usepackage{emulateapj}

\def\ref{\par\noindent\hang}

\def\spose#1{\hbox to 0pt{#1\hss}}
\let\approxlt=\lesssim
\let\approxgt=\gtrsim

\def\multleft#1{\hbox to size{\vbox {\halign {\lft{##}\cr #1}}\hfill}\par}
\def\multright#1{\hbox to size{\vbox {\halign {\rt{##}\cr #1}}\hfill}\par}

\def\degmark{^\circ}
\def\boxit#1{\vbox{\hrule\hbox{\vrule\kern3pt\vbox{\kern3pt
          #1 \kern3pt}\kern3pt\vrule}\hrule}}

\def\cm{{\rm\thinspace cm}}

\def\erg{{\rm\thinspace erg}}
\def\eV{{\rm\thinspace eV}}

\def\K{{\rm\thinspace K}}
\def\keV{{\rm\thinspace keV}}
\def\km{{\rm\thinspace km}}

\def\Mpc{{\rm\thinspace Mpc}}
\def\Msun{\hbox{$\rm\thinspace M_{\odot}$}}

\def\s{{\rm\thinspace s}}
\def\yr{{\rm\thinspace yr}}

\def\ergpcmsqps{\hbox{$\erg\cm^{-2}\s^{-1}\,$}}

\def\ergps{\hbox{$\erg\s^{-1}\,$}}

\def\kmps{\hbox{$\km\s^{-1}\,$}}

\def\Msunpyr{\hbox{$\Msun\yr^{-1}\,$}}

\def\pcmsq{\hbox{$\cm^{-2}\,$}}

\def\kmpspMpc{\hbox{$\kmps\Mpc^{-1}$}}

\makeatletter
\@ifundefined{chapter}{\def\thebibliography#1{
\section*{References}
  \list
  {\relax}{\setlength{\labelsep}{0em}
        \setlength{\itemindent}{-\bibhang}
        \setlength{\itemsep}{\parskip}
        \setlength{\parsep}{0pt}
        \setlength{\leftmargin}{\bibhang}}
    \def\newblock{\hskip .11em plus .33em minus .07em}
    \sloppy\clubpenalty4000\widowpenalty4000
    \sfcode`\.=1000\relax}}%
{\def\thebibliography#1{
  \list
  {\relax}{\setlength{\labelsep}{0em}
        \setlength{\itemindent}{-\bibhang}
        \setlength{\itemsep}{\parskip}
        \setlength{\parsep}{0pt}
        \setlength{\leftmargin}{\bibhang}}
    \def\newblock{\hskip .11em plus .33em minus .07em}
    \sloppy\clubpenalty4000\widowpenalty4000
    \sfcode`\.=1000\relax}}
 
\newlength{\bibhang}
\let\@internalcite\cite
\def\cite{\@ifstar{\citey}{\citefull}}
\def\citefull{\def\astroncite##1##2{##1\ ##2}\@internalcite}
\def\citey{\def\astroncite##1##2{##1\ (##2)}\@internalcite}
\def\citeyear{\def\astroncite##1##2{##2}\@internalcite}
\def\citename{\def\astroncite##1##2{##1}\@internalcite}
\def\@citex[#1]#2{\if@filesw\immediate\write\@auxout{\string\citation{#2}}\fi
  \def\@citea{}\@cite{\@for\@citeb:=#2\do
    {\@citea\def\@citea{; }\@ifundefined
       {b@\@citeb}{{\bf ??}\@warning
       {Citation `\@citeb' on page \thepage \space undefined}}%
{\csname b@\@citeb\endcsname}}}{#1}}
 
\def\@cite#1#2{#1\if@tempswa #2\fi} 
\def\@biblabel#1{}
 
\def\astroncite#1#2{#1\ #2}
\makeatother

\begin{document}

\title{A deep X-ray observation of NGC~4258 and its surrounding field}

\author{Christopher~S.~Reynolds\altaffilmark{1,2},
Michael~A.~Nowak\altaffilmark{1}, and Philip~R.~Maloney\altaffilmark{3}}

\altaffiltext{1}{JILA, Campus Box 440, University of Colorado,
Boulder CO~80309}

\altaffiltext{2}{Hubble Fellow}

\altaffiltext{3}{CASA, Campus Box 389, University of Colorado,
Boulder CO~80309}

\begin{abstract}
We present a deep X-ray observation of the low-luminosity active
galactic nucleus in NGC~4258 (M~106) using the {\it Advanced Satellite
for Cosmology and Astrophysics} ({\it ASCA}).  Confirming previous
results, we find that the X-ray spectrum of this source possesses
several components.  The soft X-ray spectrum ($<2\keV$) is dominated by
thermal emission from optically-thin plasma with $kT\sim 0.5\keV$.  The
hard X-ray emission is clearly due to a power-law component with photon
index $\Gamma\approx 1.8$ absorbed by a column density of $N_{\rm
H}\approx 8\times 10^{22}\pcmsq$.  The power-law is readily identified
with primary X-ray emission from the AGN central engine.  Underlying
both of these spectral components is an additional continuum which is
possibly due to thermal emission of a very hot gaseous component in the
anomalous arms and/or the integrated hard emission of X-ray binaries in
the host galaxy.  We also clearly detect a narrow iron K$\alpha$
emission line at $\sim 6.4\keV$.  No broad component is detected.  We
suggest that the bulk of this narrow line comes from the accretion disk
and, furthermore, that the power-law X-ray source which excites this
line emission (which is typically identified with a disk corona) must be
at least $\sim 100\,GM/c^2$ in extent.  {\it This is in stark contrast
to many higher-luminosity Seyfert galaxies which display a broad iron
line indicating a small ($\sim 10\,GM/c^2$) X-ray emitting region.}  It
must be stressed that this study constrains the size of the X-ray
emitting corona rather than the presence/absence of a radiatively
efficient accretion disk in the innermost regions.  If, instead, a
substantial fraction of the observed narrow line originates from
material not associated with the accretion disk, limits can be placed on
the parameter space of possible allowed relativistically broad iron
lines.  We include a discussion of various aspects of iron line
limb-darkening for highly inclined sources, including the effect of
gravitational light-bending on the apparent limb-darkening law.  By
comparing our data with previous {\it ASCA} observations, we find
marginal evidence for a change in absorbing column density through to
the central engine, and good evidence for a change in the AGN flux.  We
conclude with a brief discussion of two serendipitous sources in our
field of view; the QSO Q1218+472 and a putative $z\sim 0.3$ cluster of
galaxies.
\end{abstract}


\section{Introduction}

The nearby low-luminosity active galactic nucleus (AGN) in NGC~4258
(M~106) has become crucially important in our understanding of accreting
supermassive black holes.  Exquisite position and velocity measurements
of the H$_2$O megamasers in NGC~4258 reveal that the masing material
resides in a very thin and slightly warped disk that is in almost
perfect Keplerian motion about a central black hole with a mass of
$3.5\times 10^7\Msun$ (Nakai, Inoue \& Miyoshi 1993; Greenhill et
al. 1995a; Miyoshi et al. 1995).  Detailed studies of the maser proper
motions and centripetal accelerations confirm this interpretation
(Greenhill et al. 1995b) and allow us to measure the distance to this
source independently of the Hubble constant ($d=7.2\Mpc$; Herrnstein et
al. 1999).

Sensitive X-ray observations provide a powerful means of probing both
large scale and small scale structures within NGC~4258.  X-ray emission
was first detected in a short {\it Einstein} observatory high resolution
imager (HRI) observation (Fabbiano et al. 1992).  More sensitive soft
X-ray data collected by the {\it ROSAT} position sensitive proportional
counters (PSPC) and HRI found an extended halo of hot ($4\times 10^6\K$)
gas around NGC~4258 (Pietsch et al. 1994; Vogler \& Pietch 1999) as well
as X-ray emission associated with the well known helically twisted jets
(the anomalous arms; Pietsch et al. 1994; Cecil, Wilson \& De Pree 1995;
Vogler \& Pietch 1999).  None of these soft X-ray observations
penetrated the column of absorbing gas that obscures the AGN itself.
This had to await {\it ASCA} observations (Makishima et al. 1994;
hereafter M94). M94 found that the central X-ray source was well
described by a power-law with photon index $\Gamma\approx 1.8$, and was
absorbed by a column density of $N_{\rm H}\sim 1.5\times 10^{23}\pcmsq$.
The soft X-ray spectrum was found to be complex with components arising
from thermal plasma emission and, possibly, contributions from the
underlying X-ray binary population in NGC~4258.  A marginal detection of
an iron line was claimed with an equivalent width of $250\pm 100\eV$.

Of great importance is the opportunity that NGC~4258 gives us to study
accretion physics when the mass accretion rate is potentially very low.
The significance of this issue is highlighted when it is realized that
most of the (quiescent) supermassive black holes in the universe are
inferred to accrete matter at rates which are comparable to, or less
than, that in NGC~4258.  It was realized by several authors that when
the accretion rate is low (relative to the Eddington rate), an accretion
disk may switch into a hot, radiatively-inefficient mode (Ichimaru 1977;
Rees 1982; Narayan \& Yi 1994; Narayan \& Yi 1995).  In essence, the
plasma becomes so tenuous that the timescale for energy transfer from
the protons to the electrons (via Coulomb interactions) becomes longer
than the inflow timescale.  The energy remains as thermal energy in the
protons (which are very poor radiators) and gets advected through the
event horizon of the black hole.  These are the so-called Advection
Dominated Accretion Flows (ADAFs).  ADAFs are to be contrasted with
`standard' radiatively-efficient accretion disks in which the disk
remains cool and geometrically thin all of the way down to the black
hole (Shakura \& Sunyaev 1973; Novikov \& Thorne 1974).

X-ray observations of broad iron $K\alpha$ lines have shown that higher
luminosity systems do indeed accrete in the radiatively-efficient mode
(Tanaka et al. 1995; Fabian et al. 1995).  However, the basic nature of
the accretion disk when the mass accretion rate is low is still far from
clear.  The existence of the ADAF solution is at the mercy of poorly
known physics such as the strength of the electron-ion coupling and the
fraction of the viscous energy that is deposited into the electrons
(Quataert \& Gruzinov 1999).  Also, Blandford \& Begelman (1999) have
suggested that the ADAF solutions discussed by Narayan \& Yi (1994,
1995) are physically inconsistent and necessarily drive powerful
outflows (producing the so-called Adiabatic Inflow-Outflow Solutions;
ADIOS).  Even if ADAF-type solutions exist, it remains an open question
as to whether real disks can make a transition to this mode when the
outer regions of the disk are cold and geometrically-thin (as they
clearly are in systems such as NGC~4258).  NGC~4258 provides a
laboratory in which we can examine all of these issues.

This dichotomy in possible accretion disk physics has produced two
models for the central regions of NGC~4258.  By modeling the observed
water maser emission in this system, Neufeld \& Maloney (1995) conclude
that the accretion disk in NGC~4258 has a high efficiency ($\approxgt
10\%$), and a low accretion rate ($\dot M/\alpha \sim 10^{-4}\Msunpyr$,
where $\alpha$ is the standard viscosity parameter of accretion disk
theory).  Furthermore, the observed maser emission in part traces out a
warp in the disk.  Modeling this warp as being driven by radiation
pressure from the central X-ray source (Pringle 1996) also implies a
radiative efficiency $\sim 10\%$ (Maloney, Begelman, \& Pringle 1996).
On the other hand, Lasota et al. (1996) use ADAF models of the 2-10 keV
X-ray power law slope and continuum radio data to argue that the system
has low efficiency ($\sim 0.1\%$) and high accretion rate ($\dot
M/\alpha
\sim 10^{-2}\Msunpyr$, $\alpha \approxgt 0.3$).  While the original
model of Lasota et al. (1996) postulated a large ADAF region ($r\sim
10^5~GM/c^2$), recent radio data constrain the ADAF region to be smaller
than $r\sim 100~GM/c^2$ (Herrnstein et al. 1998).  

In this paper, we report a deep ($\sim 200$\,ksec of good exposure time)
X-ray observation of NGC~4258 with {\it ASCA} (Ohasi et al. 1996;
Makishima et al. 1996; Yamashita 1997).  We also obtained simultaneous
hard X-ray data with the {\it RXTE} satellite but, due to a gain change
prior to our observation, there are currently no robust response
matrices or background models for these particular {\it RXTE} data.  A
presentation of these {\it RXTE} data must await these developments in
calibration.  In section 2, we describe the data reduction before
discussing the X-ray spectrum of NGC~4258 in Section 3.  Section 4
focuses on the properties of the observed iron line and the implications
of this line for the nature of the accretion disk.  Section 5 presents a
brief discussion of the other interesting objects in the field of view
of the {\it ASCA} Gas Imaging Spectrometer (GIS).  Our results are
summarized in Section 6.

\section{Data reduction and the X-ray image}

\begin{figure*}[t]
\includegraphics[width=0.45\textwidth]{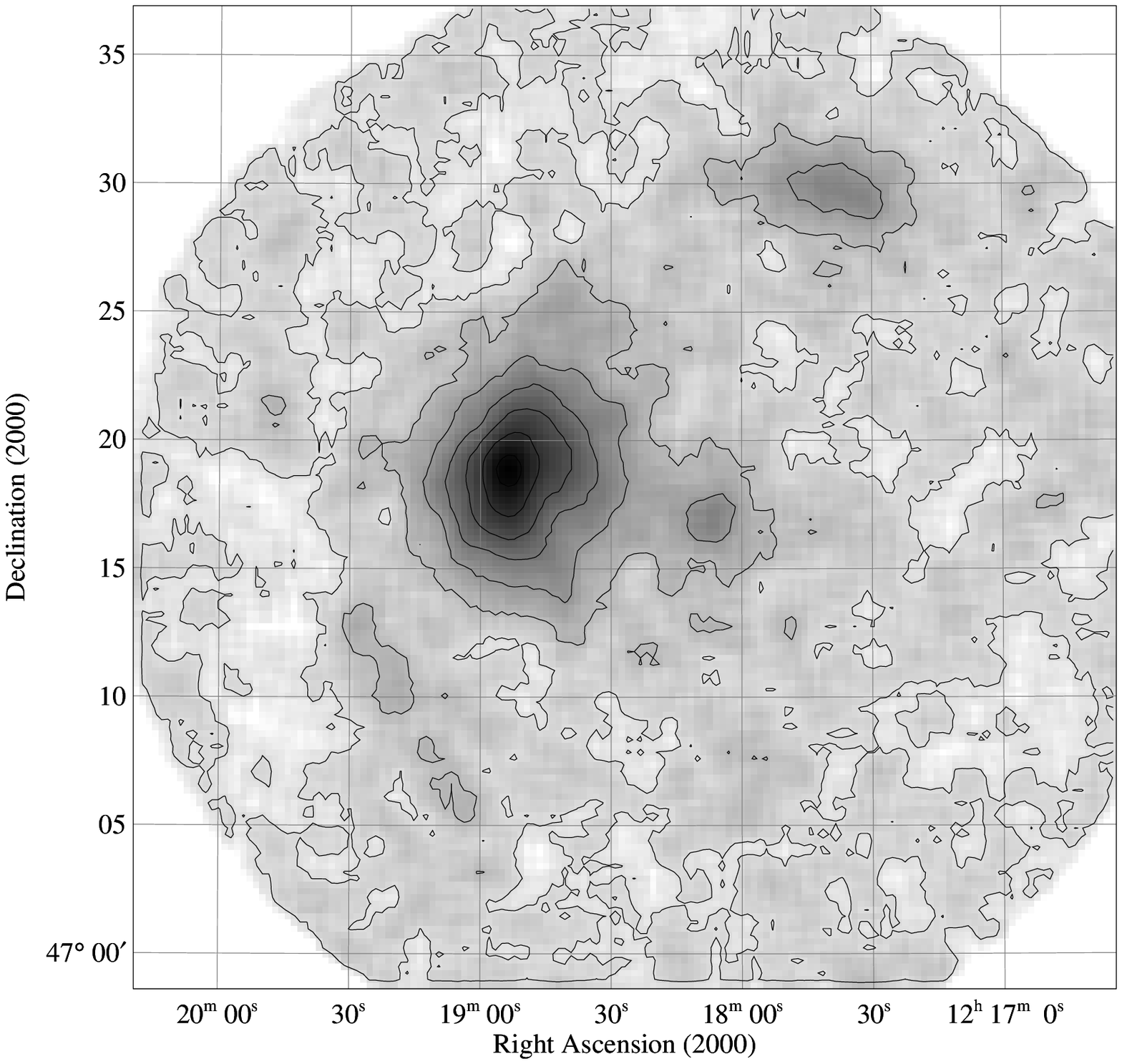}
\hspace{1cm}
\includegraphics[width=0.45\textwidth]{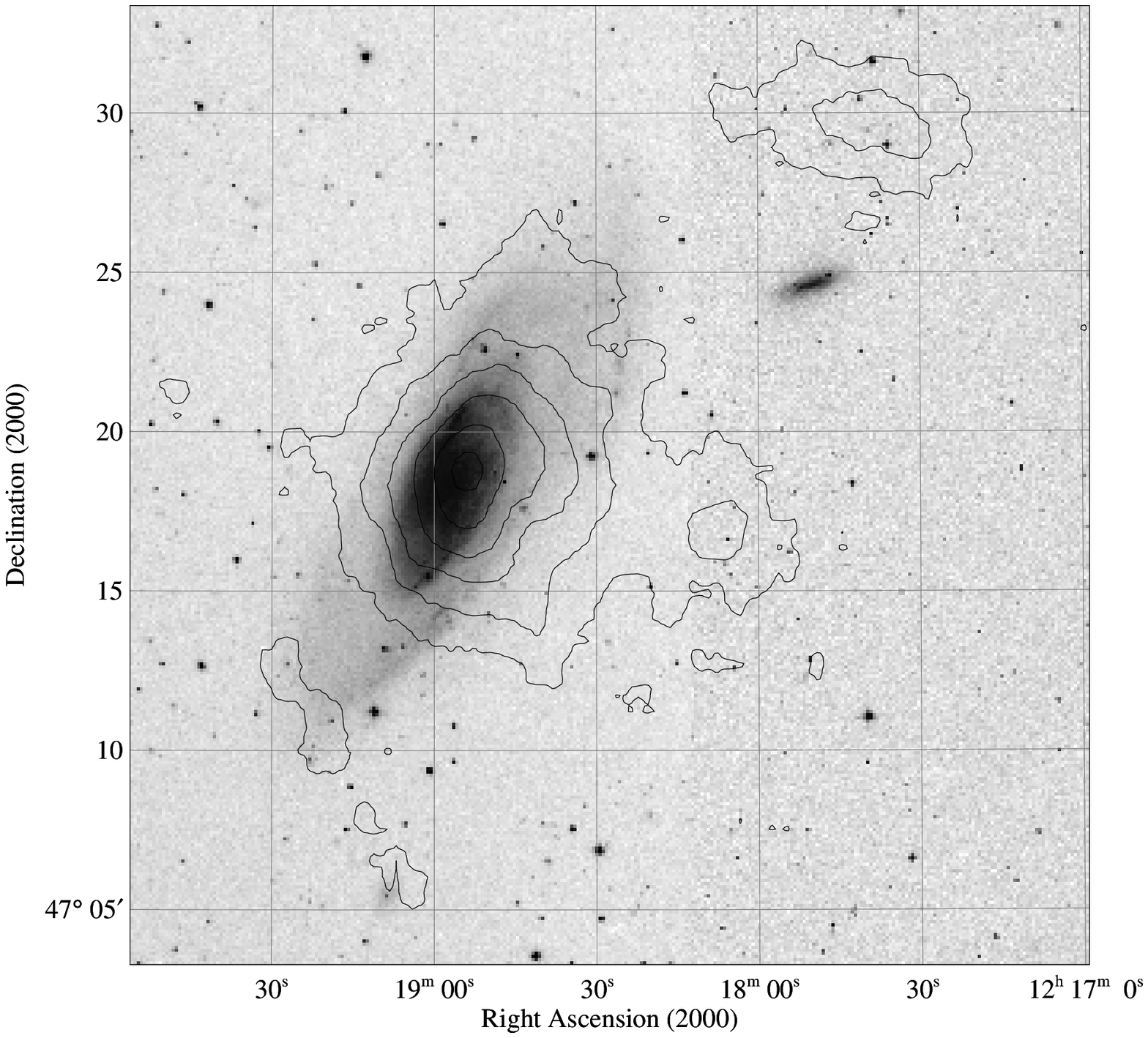}
\caption{Panel (a) shows the ASCA-GIS2 image.   These data have been
slightly smoothed for display purposes using a top-hat function with a
full-width of 1\,arcmin. Panel (b) shows X-ray contours (taken from the
smoothed GIS2 image) overlaid on the optical image from the digitized
sky survey.   The optical image was obtained from the {\sc skyview}
database situated at the NASA Goddard Space Flight Center.}
\end{figure*}

The NGC~4258 field was observed by {\it ASCA} on 1999 May 15--20.  The
SIS data were cleaned in order to remove the effects of hot and
flickering pixels and subjected to the following data-selection criteria
: i) the satellite should not be in the South Atlantic Anomaly (SAA),
ii) the object should be at least 7 degrees above the Earth's limb, iii)
the object should be at least 25 degrees above the day-time Earth limb
and iv) the local geomagnetic cut-off rigidity (COR) should be greater
than 6\,GeV/c.  We also applied a standard grade selection on SIS events
in order to further minimize particle background. The GIS data were
cleaned to remove the particle background and subjected to the following
data-selection criteria : i) the satellite should not be in the SAA, ii)
the object should be at least 7 degrees above the Earth's limb and iii)
the COR should be greater than 7\,GeV/c.  SIS and GIS data that satisfy
these criteria shall be referred to as `good' data.

After the above data selection, there are 170\,ksec of good data per SIS
detector and 185\,ksec of good data per GIS detector.  Images were then
extracted from these good data for each of the four instruments (two SIS
and two GIS).  

Fig.~1 shows the GIS2 image for this observation.  The nucleus of
NGC~4258 is the brightest X-ray source in this field.  Two other sources
are also detected: (a) one point-like source $\sim 7$\,arcmins to the
west of NGC~4258, and (b) another slightly extended source $\sim
17$\,arcmins to the north-west of NGC~4258.  Both of these sources were
clearly detected and studied with {\it ROSAT} by Pietsch et al. (1994).
Source (a) can be readily identified as the $z\approx 0.4$ quasar
Q1218+472 (Burbidge 1995; Burbidge \& Burbidge 1997).  The
identification of source (b) is less secure, but Pietsch et al. argue
that it is a background ($z\sim 0.2$) cluster of galaxies on the basis
of a possible galaxy over-density on a deep optical plate.

We have extracted spectra and lightcurves for all three sources in our
field.  Unless otherwise stated, source counts were extracted from a
circular region centered on the source with radii of 3\,arcmin and
4\,arcmin for the SIS and GIS respectively.  Background spectra were
obtained from blank regions of the same field (using the same chip in
the case of the SIS).  No temporal variability was observed in any of
these sources.  In order to facilitate $\chi^2$ spectral fitting, all
spectra were rebinned so as to contain at least 20 photons per spectral
bin.  In order to avoid poorly calibrated regions of the spectrum, the
energy ranges considered were 1.0--10\,keV for the SIS detectors, and
0.8--10\,keV for the GIS detectors.  Note that we use a lower-energy
cutoff for the SIS that is considerably higher than the `standard'
0.6\,keV cutoff in order to avoid the effects of ``residual dark current
distribution'', or RDD, which is known to plague recent {\it ASCA}
observations.

\section{A detailed X-ray study of NGC~4258}

We now discuss the X-ray properties of NGC~4258.  The two
serendipitous sources, Q1218+472 and the putative galaxy cluster, will
be addressed in Section 5.

\subsection{Fitting the soft X-ray spectrum}

\subsubsection{The inclusion of {\it ROSAT} data}

M94 showed that the soft X-ray spectrum of NGC~4258
is complex, with components arising from thermal plasma emission and,
possibly, the integrated emission of the X-ray binary population.  Since
we do not consider {\it ASCA} data below 0.8\,keV, spectrum models of
this region will be poorly constrained with many different models able
to explain the soft spectrum.  To break these degeneracies, we included
an archival {\it ROSAT} PSPC spectra in our analysis.  We chose to use
the longest single {\it ROSAT} PSPC integration of NGC~4258.  This
observation, which was performed on 1990-Jun-1, has about 25\,ksec of
good exposure time.  This dataset was obtained from the {\sc heasarc}
archive at the NASA Goddard Space Flight Center and reduced using the
{\sc ftools} routine {\sc xselect v1.4b}.  {\it ROSAT}-PSPC data were
used in the range 0.2--2\,keV.  Good agreement was obtained between the
{\it ASCA} SIS/GIS and the {\it ROSAT} PSPC in the overlap band between
1--2\,keV, thereby alleviating concerns that our analysis will be
severely affected by poor {\it ROSAT}-{\it ASCA} cross-calibration.   

\begin{table*}
{\scriptsize
\caption{Joint \textsl{ROSAT}-PSPC/\textsl{ASCA} spectral fitting results for 
NGC~4258}
\begin{center}
\begin{tabular}{clllllllllll} 
\tableline
\tableline
\noalign{\vspace*{0.7mm}}
 {Model} &  {Components} &  {$N_{\rm H}$} & {$\Gamma$} &  
 {$kT_{\rm br}$} &  {$kT_{\rm pl}$} &  {$Z_{\rm Fe,Ni}$} &  {$Z_{\rm
    light}$} &  {$E_{\rm line}$} &  {$\sigma$} & {$W_{\rm Fe}$} &
 {$\chi^2$/dof} \\ 
 {} &  {NH$_{\rm gal}$\,$\cdot$} & {$(\times 10^{22}\pcmsq)$} &  {} &
 {(keV)} &  {(keV)} &  {} &  {} & {(keV)} &   {(keV)} &  {(eV)} &  {} \\ 
 {} &  {(NH(PO)+$\ldots$)} & \\
\noalign{\vspace*{0.7mm}}
\tableline
\noalign{\vspace*{0.7mm}}
A & MEKAL$_1$ & $3.8\pm 0.3$  & $1.36\pm 0.07$ & \nodata &
$0.61\pm 0.02$ & $0.09\pm 0.01$ & $0.09\pm 0.01$ & \nodata & \nodata & \nodata 
& 1930/1506 \\
\noalign{\vspace*{0.7mm}}
B & BREMS+ & $9.7\pm 2.0$  & $1.89^{+0.39}_{-0.15}$ & $>6.2$ &
$0.36\pm 0.02$ & $0.30^{+0.08}_{-0.06}$ & $0.30^{+0.08}_{-0.06}$ & \nodata & \nodata & \nodata 
& 1627/1504 \\
& MEKAL$_1$ \\
\noalign{\vspace*{0.7mm}}
C & MEKAL$_2$ & $4.3\pm 0.3$  & $1.44\pm 0.08$ & \nodata &
$0.56\pm 0.01$ & $0.10\pm 0.01$ & $0.35\pm 0.05$ & \nodata & \nodata & \nodata
& 1784/1505 \\
\noalign{\vspace*{0.7mm}}
D & PO$_2$+ & $8.9^{+0.4}_{-0.7}$ & $1.83^{+0.06}_{-0.09}$ & 
\nodata 
& $0.34^{+0.03}_{-0.02}$ & $1.39^{+3.0}_{-0.9}$ & $1.13^{+8.0}_{-0.6}$ 
& \nodata & \nodata & \nodata & 1634/1504 \\
& MEKAL$_2$ \\
\noalign{\vspace*{0.7mm}}
E & BREMS+ & $9.5^{+2.1}_{-0.9}$ & $1.86^{+0.40}_{-0.13}$ &
$>5.0$ & $0.36^{+0.03}_{-0.02}$ & $0.31^{+0.10}_{-0.07}$ & 
$0.30\pm 0.08$ & \nodata & \nodata & \nodata & 1627/1503 \\
& MEKAL$_2$ \\
\noalign{\vspace*{0.7mm}}
F & BREMS+ & $8.2\pm 0.9$ & $1.79^{+0.31}_{-0.11}$ & 
$>4.7$ & $0.47^{+0.03}_{-0.09}$ & $0.23\pm 0.05$ & $Z_{\rm C}<2.7$ &
\nodata & \nodata & \nodata & 1557/1503 \\ 
& MEKAL$_3$  & & & & & & $Z_{\rm  N}<1.6$ \\
& & & & & &  & $Z_{\rm O}<0.14$ \\
& & & & & &  & $Z_{\rm Ne}<0.03$ \\
& & & & & &  & $Z_{\rm Na}<0.50$ \\
& & & & & &  & $Z_{\rm Mg}=0.36^{+0.11}_{-0.08}$ \\
& & & & & &  & $Z_{\rm Al}<1.3$ \\
& & & & & &  & $Z_{\rm Si}=0.7^{+1.0}_{-0.2}$ \\
& & & & & &  & $Z_{\rm  S}=1.3^{+5.0}_{-0.8}$ \\
& & & & & &  & $Z_{\rm Ar}<1.95$ \\
& & & & & &  & $Z_{\rm Ca}=5\pm 2$ \\
\noalign{\vspace*{0.7mm}}
G & BREMS+& $9.5^{+2.1}_{-0.9}$ & $1.86^{+0.40}_{-0.13}$ &
$>5.0$ & $0.36^{+0.03}_{-0.02}$ & $0.31^{+0.10}_{-0.07}$ &
$0.30\pm 0.08$ & $6.45^{+0.10}_{-0.07}$ & $<0.20$ & $107^{+42}_{-37}$ &
1609/1500 \\ 
& MEKAL$_2$+ \\
& GAU \\
\noalign{\vspace*{0.7mm}}
\tableline
\tablecomments{Model abbreviations follow: NH$_{\rm gal}$, absorption by
  the Galactic column density of $1.2\times 10^{20}\pcmsq$; NH,
  absorption by a neutral column density of $N{\rm H}$; PO, power-law
  component with photon index $\Gamma$; PO$_2$ second (unabsorbed) power
  law component with amplitude ratio $N_1/N_2=0.16$; BREMS,
  bremsstrahlung emission with temperature $kT_{\rm br}$; MEKAL$_1$,
  thermal plasma emission model (Mewe, Gronenschild \& van den Oord
  1985; Arnaud \& Rothenflug 1985; Mewe, Lemen \& van den Oord 1988;
  Kaastra 1992) with temperature $kT_{\rm pl}$ and all elemental
  abundances locked together to be a fixed fraction $Z$ of the cosmic
  value (defined by Anders \& Grevesse 1989); MEKAL$_2$, thermal plasma
  emission model in which the relative abundances of Iron and Nickel are
  fixed together (with the value $Z_{\rm Fe,Ni}$) but can vary
  independently from the relative abundance of the lighter elements
  ($Z_{\rm light}$); MEKAL$_3$, similar to MEKAL$_2$ except that the
  individual elemental abundances of the lighter elements can vary
  independently; GAU, Gaussian iron emission line component with
  rest-frame energy $E_{\rm line}$, standard deviation $\sigma$, and
  equivalent width $W_{\rm Fe}$. Errors are quoted at the 90 per cent
  level for 1 interesting parameter ($\Delta\chi^2=2.7$).}
\end{tabular}
\end{center}
}
\end{table*}

\begin{figure*}[t]
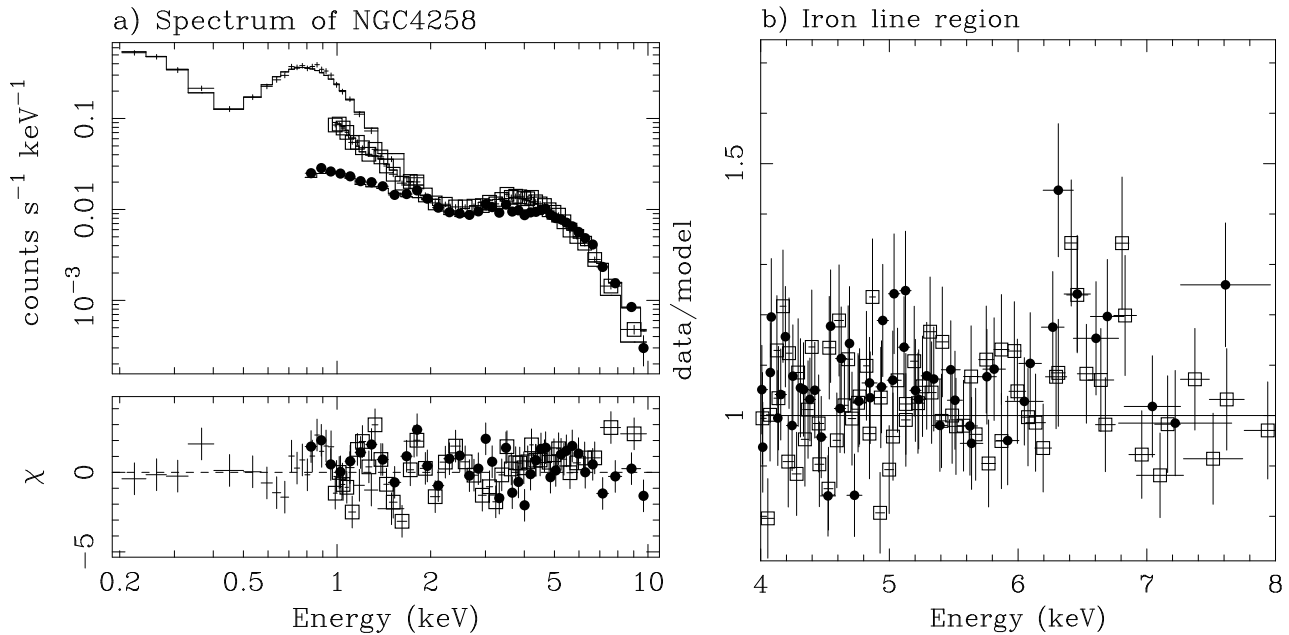

\centerline{
\includegraphics[width=0.45\textwidth,angle=270]{ngc4258_spec.ps}
\includegraphics[width=0.45\textwidth,angle=270]{ngc4258_ironline.ps}
}
\caption{Panel (a) shows the joint {\it ROSAT}/{\it ASCA} spectrum for
NGC~4258 fit with model-F from Table~1.  The {\it ROSAT} data are the
plain crosses.  For clarity, only data from the SIS0 (open squares) and
GIS2 (filled circles) instruments on board of {\it ASCA} are shown.
Panel (b) shows iron line region of the spectrum with less severe
binning, also referenced to model-F from Table~1.  The presence of a
fluorescent iron line is clear.  These are both folded spectra (in the
sense that they include the instrumental response).}
\end{figure*}

\subsubsection{Characterizing thermal plasma emission}

Table~1 details our spectral analysis of NGC~4258 (excluding our
detailed analysis of the iron line which will be addressed in Section
4).  As is evident from our spectrum and the previous work of M94, the
spectrum has both hard and soft components.  The simplest model that we
attempted to fit consists of an absorbed power-law with an additional
bremsstrahlung component, all absorbed by the Galactic column density of
$N_{\rm Gal}=1.2\times 10^{20}\pcmsq$.  This was a dreadful fit to the
spectrum, giving $\chi^2/{\rm dof}=3169/1507$.  Large residuals exist
with the model under-predicting the data in the 0.7--1.5\,keV range.
Since this is the energy range in which powerful line emission can occur
from a thermal plasma with $kT\sim 1\keV$, we next replaced the
bremsstrahlung components with the thermal plasma model {\sc mekal}
(Mewe, Gronenschild \& van den Oord 1985; Arnaud \& Rothenflug 1985;
Mewe, Lemen \& van den Oord 1988; Kaastra 1992).  Initially, we consider
a thermal plasma model in which all metals are assumed to have the same
fractional elemental abundance relative to the cosmic abundances of
Anders and Grevesse (1989).  The inclusion of the thermal plasma
component (model-A in Table~1) leads to a dramatic improvement in the
goodness of fit with $\chi^2/{\rm dof}=1925/1507$.  However, several
line-like residuals in the X-ray spectrum (including one at iron
K$\alpha$ energies), as well as a general curvature of the spectrum,
prevent this model from being an adequate fit to the data.  We now
discuss extensions of this basic spectral model which can adequately
explain the observed spectrum.

\subsubsection{An additional Bremsstrahlung-like component}

The curvature in the spectrum requires us to consider an additional
continuum component.  M94 include an additional bremsstrahlung component
in order to model the diffuse and off-nuclear emission in this object.
Including a bremsstrahlung component (model B) accounts for this
curvature and leads to a dramatic improvement in the goodness of fit
(compare models A and B; $\Delta\chi^2=298$ for 2 additional degrees of
freedom).  This additional component is required even if we allow for
abundance variations in the thermal plasma component (compare models B
and C; also see below).  For completeness, we also investigated the
possibility that this additional continuum component is a power-law that
does not suffer any intrinsic absorption (model D).  Scattering of the
AGN power-law emission around the absorbing material, or non-thermal
spatially-extended emission from the galaxy or jet would be possible
sources of such a component.  This provides a marginally worse fit than
the bremsstrahlung-based model, and implies large scattering fractions
($f_{\rm scat}=0.16$) or powerful distributed non-thermal emission
($L_{\rm X,dist}\approx 10^{40}\ergps$ in the 0.5--10\,keV range).
Hence, we think this power-law alternative to be unlikely.

\subsubsection{Constraints on the plasma abundance}

Whether this additional continuum is modeled as a bremsstrahlung or a
power-law component has little influence on the best-fit parameters for
the direct AGN component (i.e. it does not significantly affect the
photon index or normalization of the AGN power-law emission, nor the
inferred absorbing column through to the power-law source).  However, it
does effect the best-fit abundances of the soft thermal plasma component
(compare models C and B in Table~1).  In this section, we attempt to
constrain the abundances of the thermal plasma under the assumption that
this additional component has a bremsstrahlung form.  We also make the
assumption that the thermal plasma emission can be characterized by a
single temperature.  We make these assumptions here in order to be able
to make progress with these data.  Both of these assumptions may be
invalid.   We must await future high-resolution, high signal-to-noise
data in order to test and relax these assumptions through the use of
direct emission line diagnostics.

We investigated the plasma abundances in a two step process.  Firstly,
the metals were split into two classes with iron and nickel in one
class, and all of the lighter metals in the other class.  The relative
abundances were fixed within each class, but the relative abundance of
each class was allowed to vary independently.  This had no effect on the
goodness of fit, with the best-fit relative abundances of each class
being very similar (compare models B and E).  Since soft
X-ray line-like residuals still persist in these fits, the relative
abundances of all of the light elements were then allowed to vary
independently.  This leads to a further improvement in the goodness of
fit (compare models E and F; $\Delta\chi^2=70$ for 10 additional degrees
of freedom).  After fitting model F (which includes an additional
bremsstrahlung component; see below) it can be seen from Table~1 that
the abundances of C, Ne, Na, Al, and Ar are poorly constrained with only
weak upper limits possible.  This is due to the lack of strong emission
lines from these elements in the well-calibrated region of the {\it
ASCA}-SIS/GIS.  On the other hand, the abundances of Mg and Fe are well
constrained (with $Z_{\rm Mg}=0.36$ and $Z_{\rm Fe}=0.23$) due to the
detection of fairly strong emission line complexes associated with these
elements.  It is worth noting the apparent extreme overabundance of
Calcium ($Z_{\rm Ca}=5\pm 2$).  This result must be viewed with
suspicion since the dominant Calcium emission lines emerge just below
our usable {\it ASCA} band (which has a lower-energy cutoff at 0.8\,keV
for the GIS and 1.0\,keV for the SIS).  Hence, any slight {\it
ASCA}/{\it ROSAT} cross-calibration problem might be manifested as an
extreme Calcium abundance.

\subsubsection{Fluxes and luminosities}

\begin{table*}
\caption{X-ray fluxes and luminosities for NGC~4258 based on model F of
Table~1.  The bolometric luminosity of the thermal plasma emission can
been calculated under the assumption that a single temperature describes
this material.  All luminosities assume a distance of 7.2\,Mpc and have
been absorption-corrected (i.e. both Galactic and intrinsic absorption
have been removed).}
\begin{center}
\begin{tabular}{lc}\hline\hline
parameter & value \\\hline
observed 0.5--10\,keV flux & $2.1\times 10^{-12}\ergpcmsqps$ \\
observed 2--10\,keV flux & $5.8\times 10^{-12}\ergpcmsqps$ \\
power-law 0.5--10\,keV luminosity & $7.9\times 10^{40}\ergps$ \\
thermal plasma 0.5--10\,keV luminosity & $1.1\times 10^{40}\ergps$ \\
thermal plasma bolometric luminosity & $2.1\times 10^{40}\ergps$ \\
bremsstrahlung 0.5--10\,keV luminosity & $9.2\times 10^{39}\ergps$ \\\hline
\end{tabular}
\end{center}
\end{table*}

Using model F from Table~1, and assuming a distance of 7.2\,Mpc, we can
compute the absorption-corrected luminosities in each spectral
component.  The result is shown in Table~2.  It can be seen that the
power-law component dominates the energetics of the X-ray band by almost
an order of magnitude.  On energetic grounds, the thermal
plasma emission can be powered via the absorption and re-emission of
10--25\% of the hard power-law component (depending upon how low in
energy the power-law component extends).  Our data cannot probe the
nature and origin of this thermal emission beyond pointing out the basic
energetics.  However, future observations with {\it Chandra} and {\it
XMM} will be able to study the spatial distribution of the various
spectral components that we have noted.  One will then be able to
address whether the thermal emission is located in the immediate
vicinity of the AGN (as, for example, if it originated from an accretion
disk wind) or distributed on scales of 100\,pc or greater (as in the
case of a galactic superwind).

\subsection{Characterizing the iron K$\alpha$ emission line}

Residuals due to the iron line are clearly visible when models A--E are
fit to these data (see Fig.~2b).  We choose to use model D of Table~1 as
a base model for our iron line investigation.  We do not use model E
(i.e. the best fitting model from Table~1) because the excessive number
of free parameters makes the iron line error analysis unnecessarily
difficult.  We have verified that our iron line results are not affected
in any significant manner by the choice of using model D as a base model
rather than model E.  We initially fit this line by adding a Gaussian
emission feature (model F of Table~1).  The improvement in the goodness
of fit is significant at more than the 90 per cent level
($\Delta\chi^2=18$ for 3 additional degrees of freedom).  The full-width
half maximum (FWHM) of the line is constrained to be less than
$22000\kmps$, and is consistent with being zero (i.e. the line is
unresolved).  The inferred centroid energy of
$E=6.45^{+0.10}_{-0.07}\keV$ is consistent with the K$\alpha$ line
energy from iron with an ionization state of less than Fe{\sc xvii}.
The equivalent width of the line is $W_{\rm Fe}=107^{+42}_{-37}\eV$.

We discuss the origins of this emission line and the possible
implications for the nature of the accretion disk in Section 4.

\subsection{The long term variability of NGC~4258}

\begin{table*}
\caption{Historical {\it ASCA} data for NGC~4258.  Spectral parameters
shown here are derived by fitted model B of Table~1 to these data.  See
text for further details.  Errors are quoted at the 90 per cent level
for 1 interesting parameter ($\Delta\chi^2=2.7$).}
\begin{center}
\begin{tabular}{lccccc}\hline\hline
Obs. & 5-May-93 & 23-May-96 & 5-June-96 & 18-Dec-96 & 15-May-99 \\\hline
MJD & 49122 & 50226 & 50239 & 50435 & 51313 \\
SIS Exposure (1000\,s) & 39 & 23 & 31 & 26 & 170\\
$\Gamma$ & $1.78^{+0.22}_{-0.26}$ & $1.71^{+0.18}_{-0.17}$ & $1.83\pm
0.13$ & $1.87\pm 0.15$ & $1.86^{+0.40}_{-0.13}$\footnote{Bremsstrahlung
and thermal plasma parameters allowed to be free during this fit.} \\
$N_{\rm H} (10^{22}\pcmsq)$ & $13.6^{+2.1}_{-2.2}$ & $9.2\pm 0.9$ &
$8.8^{+0.7}_{-0.6}$ & $9.7\pm 0.8$ & $9.5^{+2.1}_{-0.9}$ \\
GIS2 5-10\,keV flux ($10^{-12}\ergpcmsqps$) & 5.1 & 8.3 & 8.8 & 9.5 & 4.0\\
$\chi^2$/dof & 652/511 & 639/542 & 909/728 & 828/632 & 1627/1504\\\hline
\end{tabular}
\end{center}
\end{table*}

{\it ASCA} observed NGC~4258 on 4 previous occasions, thereby allowing
us to examine the variability of this AGN on timescales of several
years.  We retrieved all of the available {\it ASCA} data on NGC~4258
from the {\sc heasarc} public database located at the NASA Goddard Space
Flight Center.  These data were reduced in the same manner as described
in Section 2, and then fitted with spectral model B from Table~1.  Since
these archival data possess significantly lower signal to noise, the
temperatures of the thermal plasma and bremsstrahlung components were
fixed to the best fit values from Table~1.  The normalization of the
bremsstrahlung component was also fixed.  Thus, this investigation is
probing changes in the AGN power-law and absorbing column density.  Note
that only GIS data were included for the 5-May-93 observation, since
NGC~4258 falls very close to the SIS chip boundaries in this
observation.

Table~3 summarized the results of this study.  The photon index of the
power-law emission is consistent with being constant over the 6 years
covered by these data.  However, there are indications of variations in
both the absorbing column density (which seems to decrease by $\sim
30\%$ between the 1993 and 1996 observations) and the 5--10\,keV
flux (which almost doubles from the 1993 to the 1996 observations, and
then drops back to the 1993 level by the time of the 1999 observation).
The flux in the 5--10\,keV range is dominated by the AGN power-law
component and is little affected by the absorbing column.

\section{Implications of the iron line in NGC~4258}

\subsection{Pure accretion disk models for the iron line}

\begin{figure*}[t]
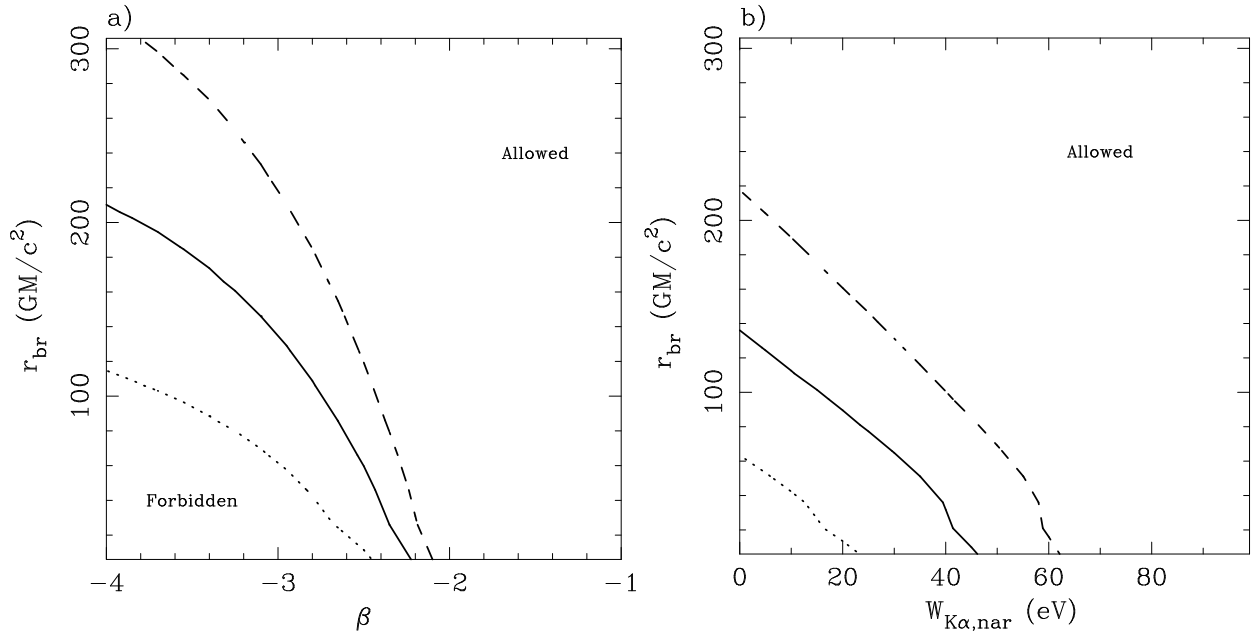

\centerline{
\includegraphics[width=0.45\textwidth,angle=270]{ngc4258_diskline_cont.ps}
\includegraphics[width=0.45\textwidth,angle=270]{diskline_inclnarr_cont.ps}
}
\caption{Panel (a) shows the confidence contours on the $\beta$-$r_{\rm
br}$ plane when the observed iron line is modelled as arising from an
accretion disk in which the line emissivity is $\epsilon=0$ for $r<r_{\rm
br}$, and $\epsilon\propto r^{\beta}$ for $r>r_{\rm br}$.  Panel (b)
shows the confidence contours on the $W_{K\alpha, nar}$-$r_{\rm br}$
plane, when it is assumed that there is also a narrow iron line from
some non-disk source with an equivalent width $W_{K\alpha, nar}$.  The
line emissivity of the disk is fixed to the canonical values of
$\beta=-3$ in this panel.  In both figures, the accretion disk is
assumed to have an inclination of $i=85\degmark$.  The confidence
contours are at the 68\% (dashed line), 90\% (solid
line) and 95\% levels (dotted line) for two interesting parameters.  In
both plots, the lower-left is the forbidden region and the upper-right
is the allowed region.}
\end{figure*}

The interest in the iron line lies in its ability to diagnose the nature
of the accretion disk.  Suppose that the only significant source of iron
K$\alpha$ line emission in NGC~4258 is the AGN accretion disk. We can
then model the observed iron line as being from the surface of the
accretion disk around a non-rotating (Schwarzschild) black hole by using
the {\sc diskline} model within the {\sc xspec} software package.  We
will make the assumption that the disk is flat inside of the masing
radii and so set the inclination of the inner disk to be $i=85\degmark$.
The outer radius of the iron line emitting region is set to $r_{\rm
out}=10^5\,GM/c^2$ (the radius of the maser disk; note that the iron
line fits are very insensitive to the actual value of $r_{\rm out}$
provided it is sufficiently large).  The line energy was fixed at the
value appropriate for K$\alpha$ emission from weakly ionized iron,
$E=6.40\keV$.  Free parameters in the fit are the inner radius of the
line emitting region $r_{\rm br}$, the index describing line emission as
a function of radius $\beta$ (where the surface emissivity
$\epsilon\propto r^{\beta}$), and the normalization of the line.
Figure~3a reports the confidence contours that result when this
accretion disk model is fit to the iron line.  It can be seen that the
narrowness of the line requires either flat emissivity as a function of
radius ($\beta>-2$) or an inner edge to the line emitting region at
greater than a hundred gravitational radii.

If the X-ray emission traces the viscous dissipation in the disk, and
the disk is geometrically-thin and radiatively-efficient so that it can
be described by a Novikov \& Thorne (1973) model, the emissivity index
should tend to $\beta=-3$ outside of the inner 20 gravitational radii or
so.  In this case, and given the assumption that the observed iron line
is from the accretion disk, we are led to the conclusion (at the 90\%
confidence level) that the line emitting region has an inner edge at
$\sim 100\,GM/c^2$.  Such an inner edge may correspond to
the point at which the disk surface becomes ionized, or where the disk
undergoes a transition to a hot (possibly advection dominated) state.
On the other hand, if the X-ray source is in the form of a
geometrically-thick corona with size $D$, the emissivity index will be
fairly flat for $r<D$, and will tend to $\beta=-3$ for $r>>D$.  In this
case (and again, with the assumption that the observed iron line is from
the accretion disk), we must conclude that the corona is large
$D\approxgt 100GM/c^2$.

\subsection{Hybrid disk/non-disk iron line models}

An alternative that we must explore is one in which a some fraction of
the observed narrow iron line is produced by distant material not
directly related to the accretion disk.  In this case, any broad
emission line from the accretion disk would be blended with this narrow
line and, possibly, buried in the noisy continuum spectrum.  To
investigate this possibility, we suppose that a narrow iron line from
some non-disk origin contributes to the observed spectrum with an
equivalent width of $W_{K\alpha, nar}$.  Fixing the line emissivity
profile of the disk to the canonical $\beta=-3$ case, Fig.~3b shows the
confidence contours on the $W_{K\alpha, nar}-r_{\rm br}$ plane.  It can
be seen that the line emission needs an inner edge or break at $r_{\rm
br}\approxgt 50GM/c^2$ unless the additional narrow line source
contributes at the level $W_{K\alpha, nar}\approxgt 40\eV$.

We cannot rule out, in any rigorous sense, the possibility that most (or
all) of the observed iron line originates from matter that is not
associated with the accretion disk.  However, simple arguments lead us
to disfavor such a scenario.  Consider iron line emission in a
geometrically-thick torus surrounding the accretion disk of NGC~4258.
An upper limit to the column density of this structure along our line of
sight to the AGN is given by the observed column density of $N_{\rm
H}\approx 10^{23}\pcmsq$.  If we suppose that this torus is in the same
plane as the accretion disk (so that we are also viewing it edge-on), it
is plausible to assume that we are looking through the
optically-thickest part of the torus.  By considering the case in which
the torus completely surrounds the AGN with uniform column density along
all radii, an upper limit to the equivalent width of the iron line is
given by:
\begin{equation}
W_{\rm Fe, max}=E_{\rm line}^2 Y_{\rm Fe}N_{\rm H}Z_{\rm
abs,Fe}\int_{0}^{\infty}\frac{\sigma(E)}{E^2}\,dE,
\end{equation}
where $E_{\rm line}=6.4\keV$ is the energy of the emission line, $Y_{\rm
Fe}=0.33$ is the fluorescent yield of the transition, $Z_{\rm
abs,Fe}\approx 4\times 10^{-5}$ is the density of iron relative to
hydrogen, and $\sigma(E)$ is the energy dependent photoionization cross
section for ionization from the $1s$ shell.  Here, we have approximated
the photon index of the ionizing AGN power-law to $\Gamma=2$ and assumed
that the central X-ray source is isotropic.  Using the photoionization
cross-sections for neutral iron of Verner \& Yakovlev (1994), this
yields
\begin{equation}
W_{\rm Fe, max}\approx 65\eV.
\end{equation}

However, there are several reasons why this upper limit would almost
certainly not be achieved.  Firstly, modeling of the accretion disk warp
strongly suggests that our line of sight intercepts the disk and that
the bulk of the column density which obscures the AGN originates in the
disk (Herrnstein, priv. communication).  Therefore, we might expect
significantly smaller column densities along lines of sight that have
smaller inclinations angles relative to the accretion disk.  Secondly,
the accretion disk may well occult half of this fluorescing cloud,
thereby reducing this prediction further.  Thus, the true iron line from
surrounding non-disk material may well be reduced from our naive
prediction by a factor of several.

A high column density torus that is misaligned with the accretion disk
so as to leave the X-ray unobscured is still a viable source for the
observed narrow iron line.  Such a torus must exist on size scales
significantly larger than the masing accretion disk, or else it would
hydrodynamically disturb the disk and lead to strong deviations from the
observed Keplerian rotation.

\subsection{Equivalent width limits on a ``Seyfert-like'' broad iron line}

\begin{figure*}[t]
\centerline{
\includegraphics[width=0.45\textwidth,angle=270]{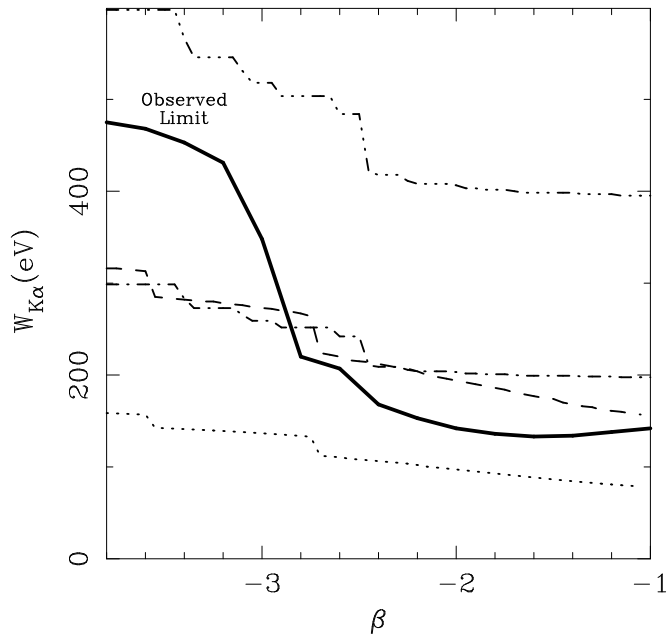}
}
\caption{Constraints on the presence of a ``Seyfert-like'' iron line in the 
case where the observed narrow line is modelled by a separate narrow
Gaussian component.  The broad line is modelled as originating in an
accretion disk around a Schwarzschild black hole, with a disk
inclination of $i=85\degmark$ and an inner line emitting radius of
$r_{\rm br}=6GM/c^2$.  The solid line shows the upper limit on the
equivalent widths a function of the emissivity index $\beta$.  The
dotted and dashed lines show the theoretical expectation, taking into
account limb-darkening and light bending effects, assuming that the iron
line has an equivalent width of $200\eV$ and $400\eV$, respectively,
when the disk is viewed face on.  The dot-dashed and dot-dot-dot-dashed
lines show the theoretical expectation when limb-darkening is absent
(see text) assuming that the iron line has an equivalent width of
$200\eV$ and $400\eV$, respectively, when the disk is viewed face on.}
\end{figure*}

Many higher luminosity Seyfert galaxies display iron lines which are so
broad that the line emitting region is thought to extend down to near
the radius of marginal stability ($6GM/c^2$ for a non-rotating black
hole).  The typical emissivity index $\beta$ lies between $-2$ and $-3$
(Nandra et al. 1997).  As shown above, if $W_{K\alpha,nar}\approxgt
40\eV$ then no such relativistically broad component is required by our
fits to NGC~4258.  However, even when all of the observed line is
assumed to arise from some other structure, we can still obtain an upper
limit to the equivalent width of any such ``Seyfert-like'' component by
fixing $r_{\rm br}=6GM/c^2$ in the spectral fitting, and including an
explicit narrow Gaussian component to fit the observed line.  The
resulting upper limit on the equivalent width of the relativistic iron
line, as a function of the assumed line emissivity index, is shown in
Fig.~4 (thick solid line).

If NGC~4258 possesses an iron line emissivity inner accretion disk
similar to higher luminosity Seyfert galaxies, one might naively expect
that limb-darkening due to absorption in the outer layers of the
accretion disk will reduce the equivalent width of any such broad iron
line to very small values (e.g. George \& Fabian 1991).  From the work
of Ghisellini, Haardt \& Matt (1994), the limb-darkening in the
plane-parallel case is well described by the expression
\begin{equation}
W_{K\alpha}(\theta)=\frac{W_{K\alpha}(\theta=0)}{\ln 2}\cos\theta\,
\ln\left(1+\frac{1}{\cos\theta} \right)
\end{equation}
where $\theta$ is the inclination angle.  Nandra et al. (1997) shows
that most Seyfert 1 galaxies have a broad iron line with $W_{K\alpha}$
in the range 200--400\,eV.  Taking these to be representative of the
face-on values, an inclination of $\theta=85\degmark$ and this
limb-darkening law reduces the expected equivalent width to
66--132\,eV.  This is below our detection threshold.

Two important effects modify this estimate.  Firstly, relativistic light
bending means that significant portions of the innermost regions of the
disk are viewed significantly more face-on than otherwise might be
thought.  The dotted and dashed lines on Fig.~4 show the effects of
light bending on the predicted broad line equivalent width (assuming a
face-on equivalent width, $W_{K\alpha}(\theta=0)$, of 200\,eV and
400\,eV respectively).  These lines have been computed using the
relativistic code presented in Reynolds et al. (1999) combined with the
above limb-darkening law.  Figure~4 shows that the broad line would not
be detectable unless $W_{K\alpha}(\theta=0)>300\eV$ and $\beta>-3$.

Secondly, the above limb-darkening law assumes that the $\tau=1$ surface
of the accretion disk is strictly planar.  It is very unlikely that this
will be the case.  MHD turbulence and violent plasma processes within
the disk corona will inevitably corrugate this surface thereby reducing
the effects of limb-darkening for highly inclined sources.  In the
extreme case, most of the matter in the outer layers of the accretion
disk may be concentrated into dense filaments or clumps (e.g. see the
MHD simulations for the `Channel solutions' case presented by Miller \&
Hawley 2000), thereby removing orientation and limb-darkening effects
almost entirely.  The dot-dashed and the dot-dot-dot-dashed lines on
Fig.~4 show the predicted equivalent width in the case of no
limb-darkening, assuming $W_{K\alpha}(\theta=0)=200\eV$ and $400\eV$,
respectively\footnote{Note that, following the convention used in {\sc
xspec}, we have defined the equivalent width with respect to the
continuum level at the energy where the line peaks in photon flux.  For
centrally concentrated disk illumination (i.e. very negative $\beta$),
the iron line peaks at $8\keV$.  Since the continuum flux is a strongly
declining function of energy, the equivalent width of such a line can
exceed the face-on value.}.  In this limiting case, we would expect to
be sensitive to such a broad line for all Seyfert-like values of $\beta$
and $W_{K\alpha}(\theta=0)$.

To summarize, these data just begin to constrain the interesting regions
of parameter space for any relativistic iron line component in NGC~4258.
There are two possible reasons for our non-detection of a broad iron
line.  Firstly, such a line might be genuinely absent thereby setting
NGC~4258 aside from its higher luminosity counterparts.  Secondly, if
NGC~4258 possesses a relativistically broad iron line with
$W_{K\alpha}(\theta=0)\approxlt 300\eV$ {\it and} limb-darkening is
important, then we would not expect to detect this line with these data.
However, we {\it can} rule out the presence of a relativistically broad
iron line with $W_{K\alpha}(\theta=0)\approxgt 300\eV$ and $\beta>-3$.
Neither do these data allow a Seyfert-like line in which limb-darkening
is unimportant.

\section{The serendipitous sources}

\subsection{Q1218+472}

\begin{figure*}[t]
\centerline{
\includegraphics[width=0.45\textwidth,angle=270]{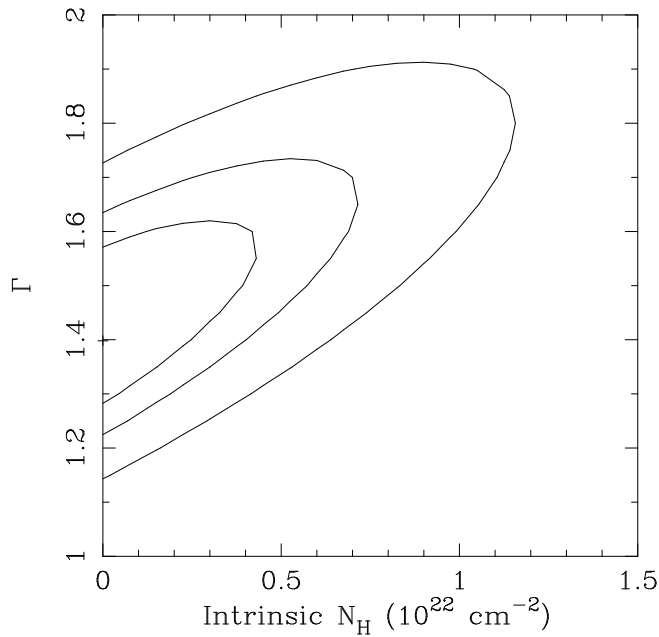}
}
\caption{Constraints on the intrinsic column density and photon
index for Q1218+472.  Shown here are the $1\sigma$, 90 per cent and 95
per cent confidence contours for two interesting parameters.}
\end{figure*}

The point source 7\,arcmins to the west of NGC~4258 is readily
identified with the $z=0.40$ quasar Q1218+472.  By choosing a slightly
smaller extraction radius than normal (2.5\,arcmins), we can isolate the
GIS counts from this source with relatively little contamination from
NGC~4258.  The resulting GIS background subtracted count rate is
$4\times 10^{-3}$\,cps per detector.  No temporal variability was
observed in the resulting light curve, although the low count rate
results in weak limits on possible variability.  This source is not
within the field of view of the SIS.

The two GIS spectra were fitted with a model consisting of a power-law
subjected to both Galactic absorption (with a column of $N_{\rm
Gal}=1.2\times 10^{20}\pcmsq$) and intrinsic absorption with the
redshift of the quasar.  The resulting best fit parameters are
$\Gamma=1.40^{+0.24}_{-0.14}$ and $N_{\rm H}<5\times 10^{21}\pcmsq$ (see
Fig.~5 for the confidence contours that result from this fit).  The
(observer frame) 2--10\,keV flux is $F_{2-10}=4.0\times
10^{-13}\ergpcmsqps$.  Assuming a redshift of $z=0.40$, a Hubble
parameter of $H_0=65\kmpspMpc$, and an acceleration parameter of
$q_0=0.5$, the rest-frame 2--10\,keV luminosity of this source is
$L_{2-10}=1.6\times 10^{44}\ergps$.  In computing the flux and
luminosity, we have applied a correction factor of 1.5 to account for
the smaller than usual extraction radius.  This correction factor was
estimated by extracting counts from other {\it ASCA} observations of
bright Seyfert galaxies with various extraction radii.

While the flux and luminosity estimate are robust, the
spectral results should be treated with caution since the GIS count
rates for this source are below the threshold at which the {\it ASCA}
Guest Observer Facility recommends spectroscopy.  Combining our X-ray
spectrum with the optical spectrum of Burbidge (1995), the spectral
index between the (rest-frame) 2500\AA\ and 2\,keV emissions is
$\alpha_{\rm ox}\approx 1.0$.  Thus, Q1218+472 is relatively X-ray
bright given its optical flux.

\subsection{The north-west source: a putative galaxy cluster}

\begin{figure*}[t]
\centerline{
\includegraphics[width=0.45\textwidth,angle=270]{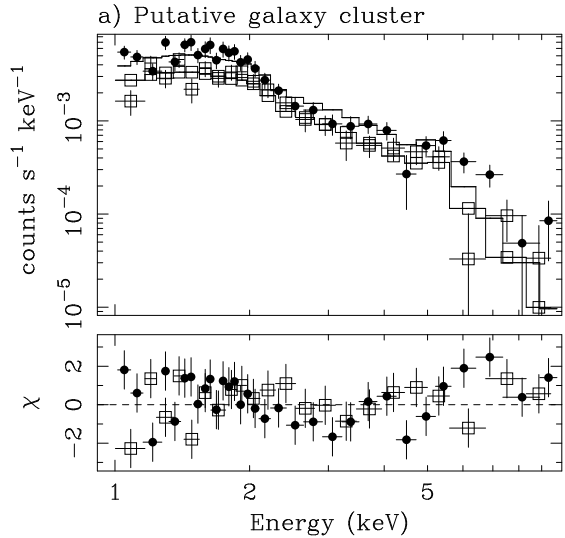}
}
\caption{GIS spectrum of the north-west object, probably a $z\sim 0.3$
cluster of galaxies.   The best fit thermal plasma model is shown (see
text for details).}
\end{figure*}

The northwest source aligns well with the extended ROSAT-PSPC source
found by Pietsch et al. (1994; hereafter P94).  P94 note a concentration
of galaxies on a deep optical plate.  By noting that the brightest
galaxy is approximately 19 magnitude in the visual band, they estimate
the redshift of the cluster to be $z\sim 0.2$.

Using standard extraction radii, we have extracted the GIS spectra and
light curves for this source (note that this source, too, fell outside
 the SIS field of view).  The GIS count rate was $7\times
10^{-3}$\,cps per detector.  Given the possible identification of this
source with a cluster of galaxies, we fit the spectrum with a thermal
plasma model (MEKAL$_1$ of Table~1) modified by Galactic absorption.  We
also allow the redshift of the cluster to be a free parameter.  The best
fit plasma temperature, relative abundance, and redshift are
$kT=5.1^{+1.3}_{-0.9}\keV$, $Z=0.49^{+0.41}_{-0.32}$ and $z=0.28\pm
0.05$ respectively.  Figure~6 shows the two GIS spectra along with this
best fit model.  It can be seen that the model is a good fit to these
data (with $\chi^2$/dof=202/217).  This X-ray determination of the
cluster redshift relies on fitting the cluster iron emission line to a
bump in the GIS spectrum at $\sim 5\keV$.  Since instrumental GIS
features are possible at these energies, this must be taken as
tentative.

The observer-frame 2--10\,keV flux of this source is $F_{2-10}=6\times
10^{-13}\ergpcmsqps$.  Assuming a cluster redshift of $z=0.28$, a Hubble
parameter of $H_0=65\kmpspMpc$, and an acceleration parameter of
$q_0=0.5$, the rest-frame 2--10\,keV luminosity of this source is
$L_{2-10}=1.4\times 10^{44}\ergps$.  These results are completely
consistent with the known relationship between cluster temperature and
X-ray luminosity (e.g. Markevitch 1998).  If, instead, this source is at
the same distance as NGC~4258, the 2--10\,keV luminosity is only
$L_{2-10}=4\times 10^{39}\ergps$.

\section{Discussion and Conclusions}

The X-ray spectrum of NGC~4258 shows 4 distinct components:
\begin{enumerate}
\item a power-law component with a 0.5--10\,keV luminosity of
$7.9\times 10^{40}\ergps$ and photon index
$\Gamma=1.79^{+0.31}_{-0.11}$.  This component is readily identified as
the primary X-ray emission from the AGN central engine itself.  This
emission is absorbed by an intrinsic column density of $N_{\rm H}=8.2\pm
0.9\times 10^{22}\pcmsq$.  Note that this absorbing column is a factor
of 1.4 smaller than that obtained by M94.  Also note that our value of
$N_{\rm H}$ is insensitive to exactly which of the acceptable spectral
models we use (cf. Gammie, Narayan \& Blandford 1999; Chary et
al. 2000).
\item thermal plasma emission from optically-thin gas at a temperature
of $kT\sim 0.5\keV$.  The total luminosity of this component is $\sim
2\times 10^{40}\ergps$, approximately 10--25\% of the power-law
luminosity (depending on how low in energy the power-law extends).
\item a bremstrahlung component with $kT>5\keV$.   This component
may represent thermal emission from a very hot gaseous component in the
anomalous arms and/or the integrated hard X-ray emission of the X-ray
binary population.
\item a narrow fluroescent K$\alpha$ emission line of cold iron with an
equivalent width of $W_{K\alpha}=107^{+42}_{-37}\eV$.
\end{enumerate}

We suggest that the bulk of the narrow iron emission line originates
from the accretion disk.  With this assumption, we have shown that the
most of the line emission must originate at relatively large radii
($r\approxgt 100GM/c^2$) in order to produce the small line width.
This, in turn, implies a large X-ray source (with a size $D\approxgt
100GM/c^2$), in contrast to the situation normally found in
higher-luminosity Seyfert galaxies (e.g. Nandra et al. 1997).  Note that
a model in which the X-ray source is small ($D\sim 10GM/c^2$) but the
inner region of the disk are too hot/ionized to produce line emission
has difficulty producing the observed equivalent width of the line.
Thus, under the assumption that the bulk of the observed iron line
originates from the accretion disk, {\it there appears to be a clear
difference in the size/structure of the X-ray source between this
low-luminosity source and higher-luminosity Seyfert galaxies.}

However, we cannot rule out the possibility that the narrow iron line
originates from some previously undetected distant matter not associated
with the accretion disk.  If the observed line {\it does} originate from
a non-disk source, our data are consistent with but do not require the
presence of a ``Seyfert-like'' relativistic broad iron line.  Even in
this case, interesting constraints can be placed on the parameter space
of possible relativistic broad iron lines.

This AGN is ripe for study with the new generation of X-ray
observatories.  The high spatial resolution of Chandra will allow us to
study large scale structure in this source directly.  We will be able
to pinpoint the location and nature of the thermal component seen in our
{\it ASCA} data.  We will also be able to spatially separate emission
from the anomalous arms, the X-ray binary population, and the AGN.  High
signal-to-noise spectroscopy with {\it XMM} (and, in the longer term,
{\it Constellation-X}) will allow the inner accretion disk to be probed.

\acknowledgements

We thank Mitch Begelman, Jim Chiang, Andy Fabian and Julian Krolik for
useful discussions throughout the course of this work.  CSR appreciates
support from Hubble Fellowship grant HF-01113.01-98A.  This grant was
awarded by the Space Telescope Institute, which is operated by the
Association of Universities for Research in Astronomy, Inc., for NASA
under contract NAS 5-26555.  We also appreciate support from NASA under
LTSA grant NAG5-6337, and the National Science Foundation under grants
AST-9529170 and AST-9876887.  PRM is supported by the NASA Astrophysical
Theory Program under grant NAG5-4061 and by NSF under grant AST-9900871.
MAN is supported by NASA under LTSA grant NAG5-3225.

\end{document}